\documentclass[twocolumn,prl,superscriptaddress]{revtex4}
\usepackage{graphicx}
\usepackage{amssymb,amsmath}
\usepackage{bm}
\usepackage{dcolumn}
\usepackage{subfigure}
\usepackage{float}
\usepackage[OT1]{fontenc} 
\usepackage{url}
\usepackage{slashed}
\usepackage{color}
\usepackage{verbatim}
\usepackage{enumitem}
\usepackage{txfonts}
\usepackage{soul,physics}
\usepackage[driverfallback=dvipdfm]{hyperref}
\hypersetup{pdfpagemode=FullScreen,colorlinks=true,breaklinks,urlcolor=blue,linkcolor=blue,citecolor=blue}
\usepackage{natbib}

\usepackage{subfigure}
\usepackage{comment}
\usepackage{hyperref}
\usepackage{amssymb}
\usepackage{bm}
\usepackage{graphicx}
\usepackage{amsmath,amssymb}
\usepackage{tikz,fp}
\usepackage{tikz-cd}
\usetikzlibrary{arrows}
\usetikzlibrary{intersections}
\usetikzlibrary{shapes.geometric}
\usetikzlibrary{decorations.pathmorphing, patterns,shapes,fixedpointarithmetic}
\usetikzlibrary{decorations.markings}

\pgfdeclarepatternformonly{south west lines}{\pgfqpoint{-0pt}{-0pt}}{\pgfqpoint{3pt}{3pt}}{\pgfqpoint{3pt}{3pt}}{
	\pgfsetlinewidth{0.4pt}
	\pgfpathmoveto{\pgfqpoint{0pt}{0pt}}
	\pgfpathlineto{\pgfqpoint{3pt}{3pt}}
	\pgfpathmoveto{\pgfqpoint{2.8pt}{-.2pt}}
	\pgfpathlineto{\pgfqpoint{3.2pt}{.2pt}}
	\pgfpathmoveto{\pgfqpoint{-.2pt}{2.8pt}}
	\pgfpathlineto{\pgfqpoint{.2pt}{3.2pt}}
	\pgfusepath{stroke}}

\tikzset{
	mid arrow/.style={postaction={decorate,decoration={
				markings,
				mark=at position .575 with {\arrow{stealth}}
	}}},
	near arrow/.style={postaction={decorate,decoration={
				markings,
				mark=at position .275 with {\arrow{stealth}}
	}}},
	far arrow/.style={postaction={decorate,decoration={
				markings,
				mark=at position .800 with {\arrow{stealth}}
	}}},
	snake arrow/.style={fixed point arithmetic, decorate, decoration={snake,amplitude=2pt, segment length=11pt},postaction={decoration={markings,mark=at position 0.625 with {\arrow{stealth}}},decorate}},
}
\tikzset{
  baseline = -0.5ex,
  wavy/.style = {
    thick,
    decorate,
    decoration={snake,amplitude=2pt,segment length=5pt}},
  sdot/.style = {
    circle,
    draw=none,
    fill=black,
    minimum size=2.5pt,
    inner sep=0pt},
  bdot/.style = {
    circle,
    draw=none,
    fill=black,
    minimum size=4pt,
    inner sep=0pt},
  svertex/.style = {
    circle,
    draw=black,
    thick,
    fill=lightgray,
    minimum size=8pt,
    inner sep=1pt},
  bvertex/.style = {
    circle,
    draw=black,
    thick,
    fill=lightgray,
    minimum size=24pt},
  bvertexsmall/.style = {
    circle,
    draw=black,
    thick,
    fill=lightgray,
    minimum size=7pt},
  bvertexnormal/.style = {
    circle,
    draw=black,
    thick,
    fill=lightgray,
    minimum size=16pt},
  dvertex/.style = {
    circle,
    draw=black,
    thick,
    fill=gray,
    minimum size=25pt}}

\begin{document}
	
	\title{Dynamical Transition of Operator Size Growth in Quantum Systems Embedded in an Environment}
	
	\author{Pengfei Zhang}
	\affiliation{Department of Physics, Fudan University, Shanghai, 200438, China}
	\affiliation{Walter Burke Institute for Theoretical Physics \& Institute for Quantum Information and Matter, California Institute of Technology, Pasadena, CA 91125, USA}
	\author{Zhenhua Yu}
	\thanks{huazhenyu2000@gmail.com}
	\affiliation{Guangdong Provincial Key Laboratory of Quantum Metrology and Sensing, School of Physics and Astronomy, Sun Yat-Sen University (Zhuhai Campus), Zhuhai 519082, China}
  \affiliation{State Key Laboratory of Optoelectronic Materials and Technologies, Sun Yat-Sen University (Guangzhou Campus), Guangzhou 510275, China}
	\date{\today}

	\begin{abstract}
	In closed generic many-body systems, unitary evolution disperses local quantum information into highly non-local objects, resulting in thermalization. Such a process is called information scrambling, whose swiftness is quantified by the operator size growth. However, for quantum systems embedded in an environment, how the couplings to the environment affect the process of information scrambling quests revelation. Here we predict a dynamical transition in quantum systems with all-to-all interactions accompanied by an environment, which separates two phases. In the dissipative phase, information scrambling halts as the operator size decays with time, while in the scrambling phase, dispersion of information persists and the operator size grows and saturates to an $O(N)$ value in the long-time limit with $N$ the number of degrees of freedom of the systems. The transition is driven by the competition between the system intrinsic and environment propelled scramblings and the environment induced dissipation. Our prediction is derived from a general argument based on epidemiological models and demonstrated analytically via solvable Brownian SYK models. We provide further evidence which suggests that the transition is generic to quantum chaotic systems when coupled to an environment. Our study sheds light on the fundamental behavior of quantum systems in the presence of an environment.

	\end{abstract}
	
	\maketitle

  \emph{ \color{blue}Introduction.--} Information scrambling emerges as a cornerstone in understanding thermalization in most closed quantum systems. 
  In such systems, the initial information, though fully preserved under the unitary evolution, is scrambled from local physical objects into those highly non-local \cite{Hayden:2007cs,Sekino:2008he,Shenker:2014cwa,Roberts:2014isa}, which gives rise to quantum thermalization of simple operators in a sufficiently long time \cite{PhysRevE.50.888,PhysRevA.43.2046}. The evolution of the operator size provides a quantitative description of the information scrambling process, and has been studied in various contexts via toy models or numerical simulations \cite{Roberts:2014isa,Nahum:2017yvy,vonKeyserlingk:2017dyr,Khemani:2017nda,Hunter-Jones:2018otn,Chen:2019klo,PhysRevLett.122.216601,Chen:2020bmq, Lucas:2020pgj,Yin:2020oze,Zhou:2021syv,Dias:2021ncd,2021PhRvR...3c2057W,sizenewpapershunyu,sizenewpaper}. In particular, closed quantum chaotic systems with all-to-all interactions are known as fast scramblers, which exhibit a strong and rapid mixing of degrees of freedom. In these systems, the operator size starts with an early-time exponential growth, and saturates as a maximally scrambled form in which all operators appear with equal probability.

Experimental study of the scrambling dynamics has been carried out in various systems \cite{2015Natur.528...77I,Li:2016xhw,Garttner:2016mqj,2019Sci...364..260B,2019arXiv190206628S,2019Natur.567...61L,2020PhRvL.124x0505J,Blok:2020may,2021PhRvA.104a2402D,2021PhRvA.104f2406D,Mi:2021gdf,Cotler:2022fin,2022PhRvA.105e2232S}. However, couplings between the systems and their environments are inevitable in experiment. Theoretical exploration of the effects of environments on the scrambling dynamics kicks off only recently and is mostly within the framework of the Lindblad equation \cite{Bhattacharya:2022gbz,2022arXiv220713603L,schuster2022operator}. Refs. \cite{2022arXiv220713603L,Bhattacharya:2022gbz} studied the information scrambling in dissipative open quantum systems from the perspective of the Krylov complexity \cite{Parker:2018yvk}, and established a relation between the complexity and the operator size distribution at least in certain cases. The authors of Ref. \cite{schuster2022operator} conducted a comprehensive study on the operator size growth in open quantum systems, nevertheless, focusing on small dissipation rates. Present knowledge in the field is rather limited.

      \begin{figure}[tb]
    \centering
    \includegraphics[width=0.75\linewidth]{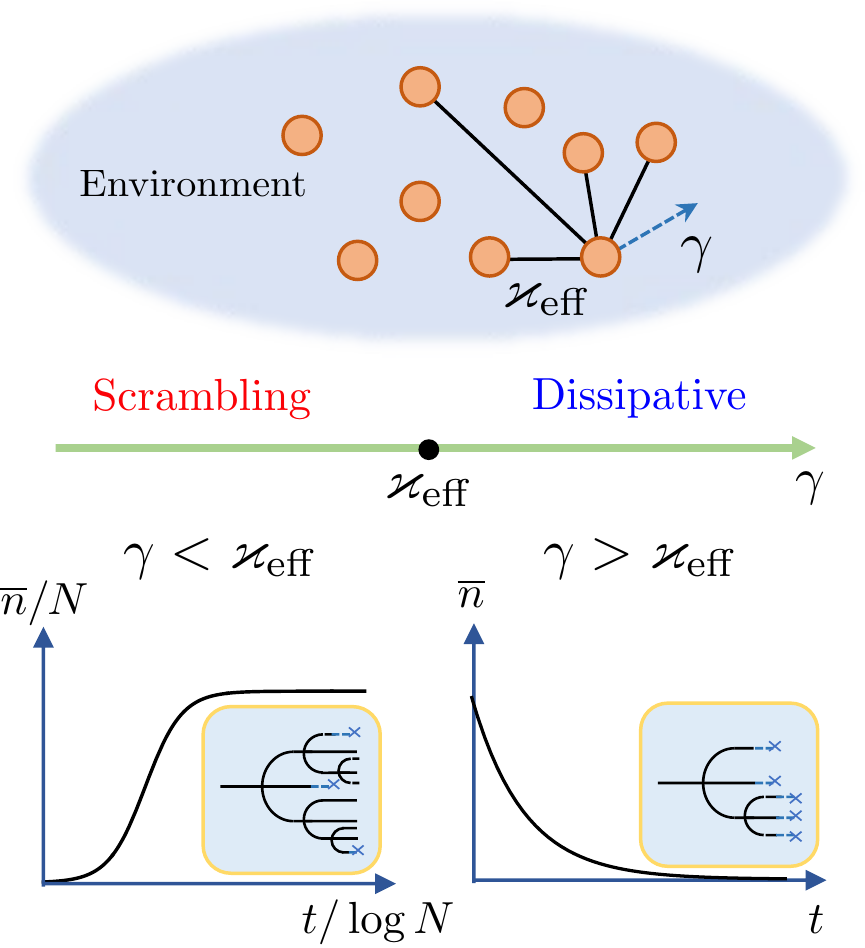}
    \caption{Schematics of the operator size growth in quantum systems with all-to-all interactions embedded in an environment. The scrambling phase and the dissipative phase are separated by the critical point when the dissipative rate $\gamma$ equals the effective net scrambling rate $\varkappa_{\rm eff}$, which is the sum of the system intrinsic scrambling rate $\varkappa$ and the environment propelled scrambling rate $\varkappa'$.}
    \label{fig:schemticas}
  \end{figure}

In this work, by a general argument based on semi-classical epidemiological models, we propose the existence of a dynamical transition of the operator size growth in quantum systems embedded in an environment with all-to-all interactions. The transition separates a scrambling phase in which information dispersion persists, and a dissipative phase in which the scrambling process halts. 
 We further demonstrate the transition by solving analytically a Brownian SYK model accompanied with an environment \cite{kitaev2015simple,maldacena2016remarks,kitaev2018soft,Saad:2018bqo,Sunderhauf:2019djv}. 
 We find that the form of system-environment couplings $H_{SE}$ is crucial in determining the effects of the environment on the system information scrambling; the part $H_{SE}^{(1)}$ linear in the system operators dissipates the process, which is quantified by a rate $\gamma$, and the rest part of $H_{SE}$ generally propels it. 
 It is the competition between the system intrinsic and environment propelled scramblings and the environment induced dissipation that gives rise to the dynamical transition. 
We argue that the transition shall be generic to quantum chaotic systems in the presence of an environment, although to experimentally observe the transition one needs to maintain access to the environment, contrary to the treatment of approximating the environment as Markovian baths \cite{Bhattacharya:2022gbz,2022arXiv220713603L,schuster2022operator}. We also discuss the implications of our results on the teleportation transition and point out possible relations between our ``environment-induced scrambling transition" to the measurement-induced entanglement phase transitions \cite{Li:2018mcv,Skinner:2018tjl,Chan:2018upn}.

  \emph{ \color{blue}Operator Size in Systems with Environments.--} We consider a class of chaotic quantum many-body systems $S$ with all-to-all interactions coupled to an environment $E$; the system $S$ and the environment $E$ consist of $N$ Majorana fermions $\chi_j$ ($j=1,2,...,N$) and $M$ Majorana fermions $\psi_a$ ($a=1,2,...,M$) respectively. We choose the normalization to be $\chi_j^2=\psi_a^2=1$. The total Hamiltonian is given by $H_{\text{tot.}}=H_S[\chi]+H_E[\psi]+H_{SE}[\chi,\psi]$. We take $M\gg N$ throughout our study \cite{Chen:2017dbb,Zhang:2019fcy,Almheiri:2019jqq}.
  
  We define the operator size distribution in the system $S$ with the environment $E$ in the following way. The time evolution of a certain system operator $O$ is determined by the Heisenberg equation as $O(t)=e^{iH_{\text{tot.}} t}Oe^{-iH_{\text{tot.}} t}$. We expand the operator $O(t)$ in terms of the complete orthonormal operator basis $\{\chi_1^{p_1}...\chi_N^{p_N}\psi_1^{q_1}...\psi_M^{q_M}\}$ with $p_j$, $q_a\in\{0,1\}$:
   \begin{equation}\label{eqn:operatorexpansion}
  O(t)=\sum_{\{p_j\}}\sum_{\{q_a\}}c_{\{p_j\},\{q_a\}}(t)\chi_1^{p_1}...\chi_N^{p_N}\psi_1^{q_1}...\psi_M^{q_M}.
  \end{equation}
The presence of $\psi_a$ in $O(t)$ is due to the system-environment coupling, distinguishing from the Lindblad formalism where the environment has been traced out \cite{wiseman2009quantum}.  
In previous studies of closed systems consisting of Majorana fermions $\chi_j$, the operator size of a string operator $\chi_{i_1}\chi_{i_2}...\chi_{i_n}$ is defined to be $n$, equal to the number of the Majorana operators \cite{roberts2018operator,Qi:2018bje,sizenewpaper}. 
To generalize to our situation, we define the size $n$ of the basis operator $\chi_1^{p_1}...\chi_N^{p_N}\psi_1^{q_1}...\psi_M^{q_M}$ by counting the number of the system Majorana operators $\chi_j$, regardless of the environment operators $\psi_a$. 
The operator size distribution $P(n,t)$ for $O(t)$ sums the weight of size $n$ basis operators in the expansion of $O(t)$: 
\begin{equation}
P(n,t)=\sum_{\{p_j\}}\delta_{n,\sum p_j}\sum_{\{q_a\}}\left|c_{\{p_j\},\{q_a\}}(t)\right|^2.
\end{equation}
  We normalize operators $O(t)$ in the way that $\text{Tr}\,O^\dagger O=2^{(M+N)/2}$; resultantly $\sum_n P(n,t)=1$. The moments of the operator size are given by $\overline{n^k}=\sum_n n^k P(n,t)$.

\emph{ \color{blue}Epidemiological models.--} The results of operator size growth in closed systems with all-to-all interactions have been interpreted intuitively by epidemiological models \cite{roberts2018operator,Qi:2018bje}, in which the number of elementary Majorana operators $\chi_j$ is an analog of the number of patients.
 For a Hamiltonian $H_S$ with $q$-body interactions of strength $J$, the commutator $i[H_S,O]$ in the Heisenberg equation of $O(t)$ brings about the infection of $(q-2)$ previously unexposed people at a rate $\sim J$. In the early-time regime $t\sim O(N^0)$, the epidemiological model thus leads to the average operator size $\overline n$ satisfying $d\overline{n}/dt=J (q-2)\overline{n}\equiv \varkappa\overline{n}$, agreeing with diagrammatic calculations \cite{roberts2018operator,Qi:2018bje}. Here $\varkappa$ is the system intrinsic quantum Lyapunov exponent. 
 
To generalize the epidemiological model to our situation, a special case of broad interest is that the system-environment coupling $H_{SE}$ takes the bilinear form $i\sum_{ja}\gamma_{ja}\chi_j \psi_a$, applicable to ubiquitous situations where there is single fermion hopping between systems and their environments. We go beyond and consider the general form of system-environment couplings
\begin{equation}\label{supp:generalHint}
H_{SE}={\sum_{n,m}}'\sum_{i_1<...<i_n}\sum_{a_1<...<a_m}i^{\frac{n+m}{2}}V_{\{i_p\};\{a_q\}}\chi_{i_1}...\chi_{i_n}\psi_{a_1}...\psi_{a_m},
\end{equation}
where the primed sum requires $n+m$ being even. We further divide $H_{SE}$ into the part $H_{SE}^{(1)}$ linear in $\chi_j$, the part $H_{SE}^{(2)}$ quadratic in $\chi_j$ and the rest part $H_{SE}'$.
The environment affects the operator size via $i[H_{SE},O]$: $H_{SE}^{(1)}$ can replace one $\chi_i$ by one $\psi_a$ \footnote{The contribution from the process that some environment Majorana fermion operator is replaced by system Majorana fermion operators vanishes when taking $M\rightarrow\infty$ with proper scaling of $M$ for the system-environment coupling constants \cite{Chen:2017dbb}.}, resulting in a decrease of the operator size, while $H_{SE}^{(2)}$ does not change the operator size, and $H_{SE}'$ leads to its increase, propelling the scrambling. 
Conrrespondingly, the generalized epidemiological model yields in the early-time regime 
 \begin{equation}\label{eqn:guess}
  \frac{d\overline{n}}{dt}=\varkappa\overline{n}-\gamma\overline{n}+\varkappa'\overline{n},
  \end{equation}
  with the dissipation rate $\gamma$ and the propelled scrambling rate $\varkappa'$ pertaining to the effects of $H_{SE}^{(1)}$ and $H_{SE}'$ respectively.
The solution $\overline{n}\sim e^{(\varkappa_{\rm eff}-\gamma)t}$, with the effective net scrambling rate $\varkappa_{\rm eff}\equiv\varkappa+\varkappa'$, indicates two distinct dynamical phases separated by the critical point at $\gamma_c=\varkappa_{\rm eff}$. For $\gamma<\gamma_c$, the average operator size increases exponentially as its counterparts in the closed systems; we take the exponential increase as the defining feature of the scrambling phase. The exponential growth of $\overline{n}$ is expected to terminate at an enhanced scrambling time $t_s\sim (\gamma_c-\gamma)^{-1}\ln N$, where additional information is required to analyze the full operator dynamics. For $\gamma>\gamma_c$, the average operator size decreases to exponentially small value after time $(\gamma-\gamma_c)^{-1}$, which suggests only environment operators remain in the expansion \eqref{eqn:operatorexpansion}; we call this the dissipative phase. At the critical point $\gamma_c=\varkappa_{\rm eff}$, $\overline{n}$ is expected not to change with time. However, as we show below through an explicit model calculation, the operator size distribution $P(n,t)$ exhibits a non-trivial power-law decrease in time $t$ for any $n\geq 1$. Schematics for the operator size growth in different phases are presented in FIG. (\ref{fig:schemticas}).
  \vspace{5pt}

  \emph{ \color{blue}Solvable SYK Models.--} We demonstrate the dynamical transition via solvable Brownian SYK models \cite{Saad:2018bqo,Sunderhauf:2019djv}, which are closely related to Brownian circuits in spin models \cite{Zhou_Brownian_2019}. 
  The specific class of the Hamiltonians we consider have the form
  \begin{equation}\label{eqn:H1}
  H(t)=\sum_{i_1<...<i_q}i^{\frac{q}{2}}J_{\{i_p\}}(t)\chi_{i_1}\chi_{i_2}...\chi_{i_q}+H_{SE}(t)+H_E
    \end{equation}
  for even $q$.
  Here $H_{SE}(t)$ takes the form of (\ref{supp:generalHint}) with time dependent coupling constants $V_{\{i_p\};\{a_q\}}(t)$. 
The coupling constants $J_{\{i_p\}}(t)$ and $V_{\{i_p\};\{a_q\}}(t)$ with different indices are independent Brownian variables, satisfying 
  \begin{equation}
  \begin{aligned}
  &\overline{J_{\{i_p\}}(t_1)J_{\{i_p\}}(t_2)}={(q-1)!J}\delta(t_1-t_2)/{N^{q-1}},\\
  &\overline{
V_{\{i_p\};\{a_q\}}(t_1)V_{\{i_p\};\{a_q\}}(t_2)}=(n-1)!m!V_{nm}\delta(t_1-t_2)/N^{n-1}M^m.
  \end{aligned}
  \end{equation}
We take the limit of $M\gg N\gg 1$ and work out the Brownian SYK models by the $1/N$ expansion \cite{maldacena2016remarks}. Note the exact form of $H_E$ is irrelevant in the following calculation since only $\langle\psi_a(t)\psi_a(t)\rangle$ is need, which is unity.

To study the operator size growth, we generalize a method introduced in \cite{Qi:2018bje}: We construct auxiliary Majorana fermions $\tilde \chi_j$ and $\tilde \psi_a$, which are paired up with $\chi_j$ and $\psi_a$ to form the complex fermions $c_j=(\chi_j+i\tilde \chi_j)/2$ and $c_a^E=(\psi_a+i\tilde \psi_a)/2$ respectively. 
We introduce a normalized EPR state $|\text{EPR}\rangle$ which satisfies $c_j|\text{EPR}\rangle=c_a^E|\text{EPR}\rangle=0$. Consequently, when we apply $O(t)$ to $|\text{EPR}\rangle$, the complex fermion number can never decrease; each Majorana operator $\chi_j/\psi_a$, when acting on $|\text{EPR}\rangle$, creates one excitation $c^\dagger_j/c^{E,\dagger}_a$. Thus the operator size distribution can be calculated via $P(n,t)=\langle O(t)^\dagger\Pi(n)O(t)\rangle$, where $ \Pi(n)$ is the projection operator into the subspace with $n$ excitations of  $c^\dagger_j$ while leaving the number of $c^{E,\dagger}_a$ excitations arbitrary. The expectation values $\langle\dots\rangle$ are taken with respect to $|\text{EPR}\rangle$.

It is convenient to work with the generating function $z(\mu,t)=\sum e^{-\mu n} P(n,t)$, which takes the form 
\begin{equation}
z(\mu,t)=\langle O(t)^\dagger e^{-\mu \sum_j c^\dagger_j c_j}O(t)\rangle.
\end{equation}
  In particular $\overline{n}=-\partial_\mu z(0,t)=\frac14\sum_j\left\langle\left|[O(t),\chi_j]\right|^2\right\rangle$, where the commutator shall be replaced by the anti-commutator if $O$ is fermionic.
  In the following discussions, we focus on the simplest non-trivial case $O=\chi_1$. We work out the retarded Green's functions $G^R(t)\equiv -i\theta(t)\left<\{\chi_i(t),\chi_i(0)\}\right>$ on the Keldysh contour as in \cite{Zhang:2020jhn} and find $G^R(\omega)^{-1}=\omega/2-\Sigma^R$ and $\Sigma^R=-i \left(J+{\sum_{n}}U_{n}\right)\equiv -i\Gamma/4$ with $U_{n}\equiv{\sum'_{m}}V_{nm}$;
the Fourier transform gives $G^R(t)=-2i\theta(t)e^{-\Gamma t/2}$. 
  The generating function $z(\mu,t)$ turns into a two-point function on the double Keldysh contour \cite{Aleiner:2016eni} with a perturbation source:
 $ \Delta I=\frac{1}{4}(1-e^{-\mu})
  \sum_j\left[\chi_j^{u_2}(0)-\chi_j^{d_2}(0)\right]\left[\chi_j^{d_1}(0)-\chi_j^{u_1}(0)\right]$.
  Here $\chi_j^{u_k/d_k}(t)$ labels the Majorana fields on the double Keldysh contour \cite{Zhang:2020jhn}: $u/d$ denotes the forward/backward evolution and $k=1$ or $2$ labels the two worlds. The factor $(1-e^{-\mu})$ appears because $\langle \bar{\phi}|e^{-\mu c^\dagger c}|\phi\rangle=e^{-e^{-\mu}\bar{\phi}\phi}$ for coherent states $|\phi\rangle$. 
  
    	\begin{figure}[tb]
		\centering
		\includegraphics[width=0.98\linewidth]{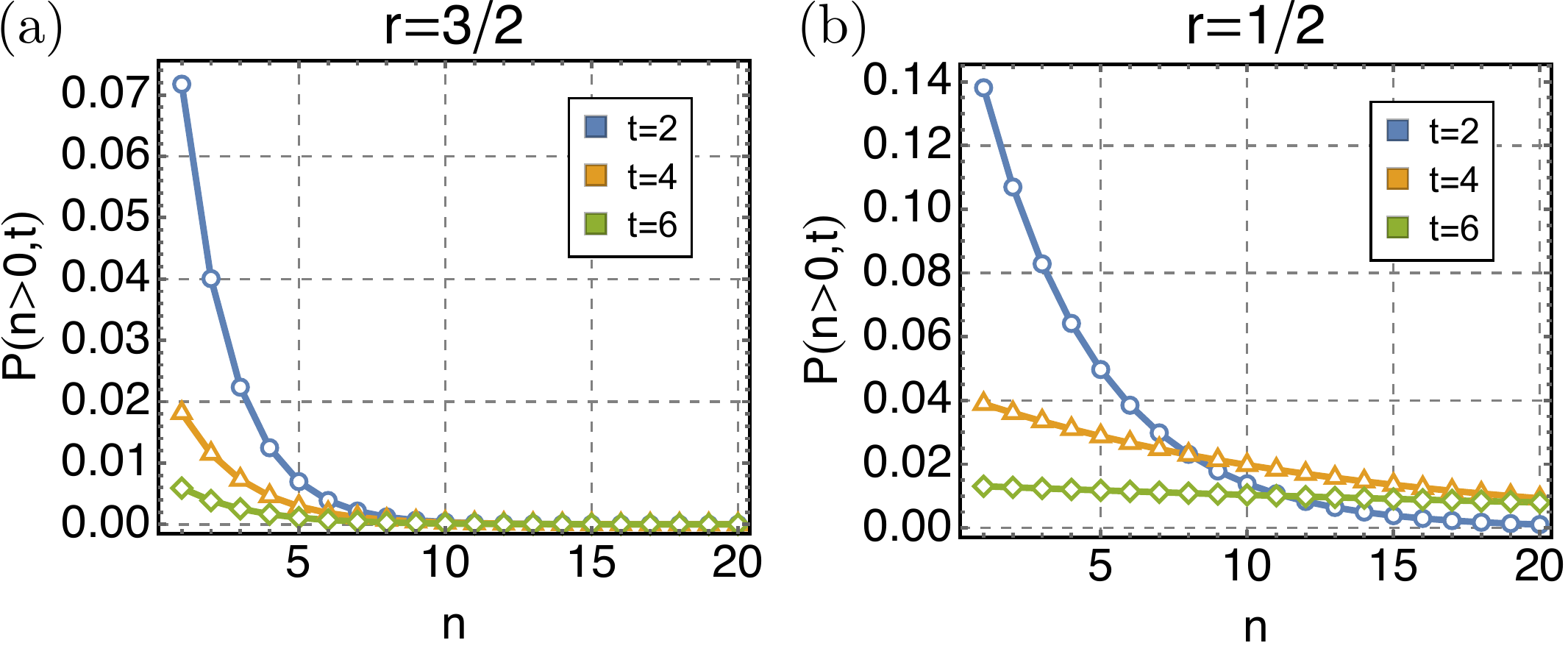}
		\caption{The plot for the early-time operator size distribution $P(n>0,t)$ of the model (\ref{eqn:H1}) with $J=0$ and $U_{n}=0$ for $n\neq1,3$ in different phases: (a) In the dissipative phase with $r\equiv \gamma/\varkappa_{\rm eff}=3/2$, the operator weight for $n\geq 1$ decays to zero quickly. (b) In the scrambling phase with $r\equiv \gamma/\varkappa_{\rm eff}=1/2$, the operator weight for $n\geq 1$ remains finite since $P(0,t)=r$ for $t\gg 1$. The symbols are generated from (\ref{eqn:solP}). The linking lines are guide for the eye. We take the units such that $\varkappa_{\rm eff}=1$.}
		\label{fig:Early-time_size}
	\end{figure}
  
   \emph{ \color{blue}Early-time Operator Dynamics.--}
   As our calculations are carried out via the $1/N$ expansion, the early-time regime $t\sim O(N^0)$ and the late-time regime $t\sim O(\ln N)$ are both defined by how time $t$ (with proper units taken) compares with orders of $N$ in the limit $N\to\infty$. In the early-time regime $t\sim O(N^0)$, we can derive the equation of $z(\mu,t)$ using the saddle-point equation \cite{SM}
   \begin{equation}\label{eqn:simple_eqn}
   \frac{dz}{dt}=4J(z^{q-1}-z)+4{\sum_{n}}U_{n}(z^{n-1}-z), \ \ \ \ \ z(\mu,0^+)=e^{-\mu}.
   \end{equation}
From \eqref{eqn:simple_eqn}, taking derivative with respective to $\mu$ and setting $\mu=0$, we find $d\overline{n}/dt=4\left[J(q-2)+{\sum_{n}}U_{n}(n-2)\right]\overline{n}$
, which matches  \eqref{eqn:guess} with identifications $\varkappa=4J(q-2)$, $\gamma =4U_{1}$ and $\varkappa'=4\sum_{n\geq 3}'(n-2)U_{n}$. This agreement suggests that the solvable SYK models we constructed show the dynamical phase transition between the scrambling phase and the dissipative phase. 
   
   To fully characterize the operator dynamics in the early-time regime, we need to solve the generating function via  \eqref{eqn:simple_eqn}. Let us focus on the case of $J=0$ and $U_{n}=0$ for $n\neq1,3$ where the evolution of the operator size distribution can be computed in closed-form. 
   Though in this particular case the system intrinsic scrambling rate $\varkappa$ is zero, and there is only scrambling propelled by the environment, and $\varkappa_{\rm eff}=\varkappa'=4U_3$. Given \eqref{eqn:simple_eqn} being the basis for further calculations, the exact origins of scrambling are expected to make no essential difference \cite{SM}. We introduce $r\equiv \gamma/\varkappa_{\rm eff}$, and set $\varkappa_{\rm eff}=1$ for conciseness. 
      
   Assuming $r\neq 1$, integrating \eqref{eqn:simple_eqn} yields   \begin{equation}\label{eqn:solz}
    z(\mu,t)=1-\frac{(r-1)(1-e^{-\mu})}{e^{(r-1)t}(r-e^{-\mu})-(1-e^{-\mu})}.
   \end{equation}
   Expanding in terms of $e^{-\mu}$, we find   
   \begin{equation}\label{eqn:solP}
   P(n,t)=\begin{cases}
      1-(1-r)\frac{e^{(1-r)t}}{e^{(1-r)t}-r}, & (n=0)\\
      (1-r)^2\frac{e^{(1-r)t}[e^{(1-r)t}-1]^{n-1}}{[e^{(1-r)t}-r]^{n+1}}. & (n\geq 1)
    \end{cases} 
   \end{equation}
In the dissipative phase with $r>1$, \eqref{eqn:solP} shows the operator weight is localized near $n=0$, with an exponentially decaying tail $P(n,t)\sim e^{-n/\xi(t)}$ and $\xi(t)=1/\ln[(r-e^{(1-r)t})/(1-e^{(1-r)t})]$. This feature is illustrated in FIG. \ref{fig:Early-time_size} (a) for $r=3/2$. Eventually, only environment operators survive in the expansion of $\chi_1(t)$. 
On the other hand, in the scrambling phase $r<1$, $P(0,t)\approx r$ for $(1-r)t\gg1$; since the distribution is normalized, the sum of the operator weight at $n\geq 1$ is always finite. Furthermore, \eqref{eqn:solP} shows that the operators with size $1 \leq n\lesssim e^{(1-r)t}$ are occupied by the same order, as illustrated in FIG. \ref{fig:Early-time_size} (b).
Contrary to closed systems \cite{Aleiner:2016eni}, the out-of-time-order correlation in the system $S$ would decay to instead a finite residual due to the coupling induced by the environment $E$, which is reflected by the nonzero value of $P(0,t)$.
Extrapolating from the early-time regime to $t\sim O(\ln N)$, a typical operator would have size $\sim N$ in the scrambling phase, and the saddle-point description \eqref{eqn:simple_eqn} is no longer sufficient \cite{gu2022two}. It is necessary to go beyond and employ the full scramblon effective theory \cite{gu2022two,sizenewpaper} to compute the operator size distribution \footnote{An alternative approach can be applied to the Brownian SYK model \cite{sizenewpapershunyu}.}. 

At the critical point with $r=1$, we have
   \begin{equation}
   \begin{aligned}
   z(\mu,t)&=1-\frac{1-e^{-\mu}}{1+t(1-e^{-\mu})},\\
   P(n,t)&=\begin{cases}
      1-\frac{1}{1+t}, & (n=0)\\
      \frac{t^{n-1}}{(1+t)^{n+1}}. & (n\geq 1)
    \end{cases}
   \end{aligned}
   \end{equation}
  The operators with $1 \leq n\lesssim t$ are occupied by the same order. However, for $t\gg 1$, $P(0,t)\to1$; all the information dives into the environment $E$. Consequently, there are no non-trivial dynamics in the late-time regime. No matter in the dissipative or scrambling phases, or at the critical point, the monotonic increase of $P(n=0,t)$ with time renders the picture that the environment operators can be viewed an attractor, continuously absorbing the weight from the system operators in the early-time regime.

      \begin{figure}[tb]
    \centering
    \includegraphics[width=0.8\linewidth]{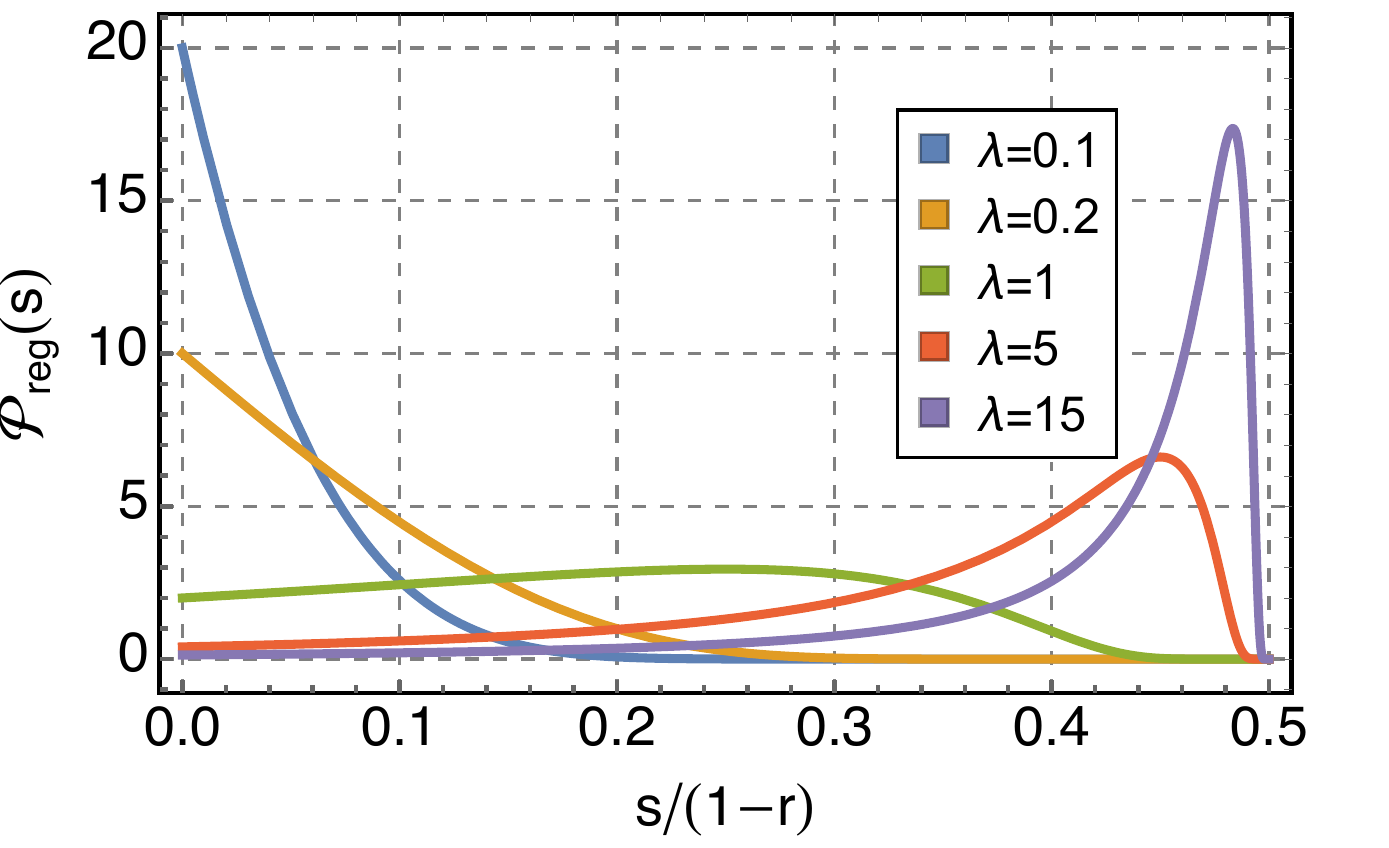}
    \caption{The operator size distribution $\mathcal{P}_{\text{reg}}(s,t)$ of the model (\ref{eqn:H1}) with $J=0$ and $U_{n}=0$ for $n\neq1,3$ in the scrambling phase ($r\equiv \gamma/\varkappa_{\rm eff}<1$) in the late time regime $t\sim O(\ln N)$. The curves are generated from (\ref{eqn:reslate}), which is obtained in the large $N$ limit with fixed $\lambda$. For the same value of $\lambda$, $\mathcal{P}_{\text{reg}}(s,t)$ for different $r$ collapse to a universal function of $s/(1-r)$.}
    \label{fig:Late-time_size}
  \end{figure}
   \emph{ \color{blue}Scrambling at the Late Time.--} We now study the late-time operator size distribution in the scrambling phase using the scramblon effective theory \cite{gu2022two}. Since, in the late-time regime $t\sim O(\ln N$), a typical operator has size $n\propto N$, we consider a continuum limit with $s=n/N\in[0,1]$ and $\mathcal{P}(s,t)\equiv NP(n,t)$ as in the closed quantum systems \cite{sizenewpaper}. We have $\int_0^1 ds~\mathcal{P}(s,t)=1$, and the generating function $  \mathcal{S}(\nu,t)\equiv\int_0^1 ds~e^{-\nu s}\mathcal{P}(s,t)=z({\nu}/{N},t)$.
   Our task is to compute $\mathcal{S}(\nu,t)$ for $t\sim t_s$ by making use of $z(\mu,t)$ obtained for $t\sim O(N^0)$. 
   
   The crucial observation is that important corrections to the saddle-point solution are only induced by the interaction between fermions that are mediated by scramblons \cite{gu2022two}. The scramblons are a collective mode having a soft action such that their propagator increases exponentially with time \cite{Stanford:2021bhl}; this softness reflects intrinsic inter-world correlations on the double Keldysh contour \cite{aleiner2016microscopic}. In our case, the expressions of the scramblon propagator and the scattering vertices between fermions and scramblons can be determined using \eqref{eqn:solz} for $z(\mu,t)$. The late-time operator-size distribution $\mathcal{S}(\nu,t)$ is then computed by summing up scramblon diagrams to the leading order in $1/N$ as in \cite{sizenewpaper}. The final result is given by \cite{SM}
   \begin{equation}\label{eqn:reslate}
   \begin{aligned}
   \mathcal{P}(s,t)=&r \delta (s)+\mathcal{P}_\text{reg}(s,t)\\
   \mathcal{P}_\text{reg}(s,t)=&\theta\left(1-\frac{2s}{1-r}\right)\frac{2 (r-1)^2 e^{\frac{2 s}{\lambda  (r+2 s-1)}}}{\lambda  (r+2 s-1)^2},
   \end{aligned}
   \end{equation}
   where $\lambda=e^{(1-r)t}/C$ with $C=N(1-r)^2/2$ is the propagator of the scramblons. The singular part $\delta (s)$ is consistent with the saturation value of $P(0,\infty)$ derived in \eqref{eqn:solP}. This part means that in the late-time regime, the weight of pure environment operators does not change. This is because the typical operators appearing in the expansion contain $O(N)$ system fermions, which have negligible coupling to the $n=0$ operators. The regular part $\mathcal{P}_\text{reg}(s,t)$ describes operators scrambled into the large size, with a total weight $(1-r)$.  Interestingly, when $\lambda$ is fixed, $\mathcal{P}_\text{reg}(s,t)$ depends on $s$ and the parameter $r$ only through the combination $s/(1-r)$. Note in the late-time regime $t\sim O(\ln N)$, $\lambda\sim O(N^0)$. For closed quantum systems, the operator becomes maximally scrambled in the long-time limit ($t\rightarrow \infty$), which corresponds to a delta peak at $s=1/2$ \cite{sizenewpaper}. However, in our situation, we see that via the linear part $H_{SE}^{(1)}$ of system-environment couplings, the environment exerts a ``drag'' and renders the analog of the maximally scrambled operator having a typical size $s=(1-r)/2$. A plot of $\mathcal{P}_\text{reg}(s,t)$ is presented in FIG. \ref{fig:Late-time_size}. Using \eqref{eqn:reslate}, we could also compute moments of the operator size $\overline{n^k}$. As an example, we find the average operator size reads $\overline{n}=N\overline{s}=N\frac{(1-r)^2 }{2}{\left[1 -\frac{1}{\lambda}e^{\frac{1}{\lambda }} \Gamma \left(0,\frac{1}{\lambda }\right)\right]}$, which increases monotonically to the saturation value $N\frac{(1-r)^2 }{2}$. 
   
  \emph{ \color{blue}Discussions.--} 
In this work, based on the epidemiological models and the solvable Brownian SYK models, we show that a dynamical transition of the operator size growth exists in quantum systems with all-to-all interactions in the presence of an environment. A crucial ingredient for the transition is $H_{SE}$ consisting of a part $H_{SE}^{(1)}$ linear in the system operators $\chi_j$; the transition is shown robust against the other couplings. Moreover, extended calculations of a Brownian SYK chain and a static SYK model provide evidence which suggests that the transition is generic to quantum chaotic systems when coupled to an environment \cite{SM}.

These results imply that in the situation that one has access to not only the system but also the environment, including the capacity to prepare the environment together with its auxiliary right system in EPR pairs initially, 
in the scrambling phase, 
teleportation is still possible in non-holographic set-ups \cite{Nezami:2021yaq,PhysRevX.12.031013,Yoshida:2017non,telenewpaper}, corresponding to our Brownian SYK models, and maybe even in holographic ones \cite{Gao:2016bin,Maldacena:2017axo,Susskind:2017nto,Gao:2018yzk,Brown:2019hmk,Gao:2019nyj} based on evidence from the static SYK model \cite{SM}.
Our ``environment-induced scrambling phase transition'' shall have a close relation with the celebrated ``measurement-induced entanglement phase transition'' \cite{Li:2018mcv,Skinner:2018tjl,Chan:2018upn}. 
The volume-law phase in the presence of measurement is known to be protected from local quantum errors \cite{Gullans:2019zdf,Choi:2019nhg,Fan:2020sau}, so is expected the scrambling phase in our situation. Furthermore, the averaged operator size $\overline n$ conveys information similar as the entropy of a system \cite{Hosur:2015ylk,2017SciBu..62..707F}. We postpone a detailed study to future works.

\vspace{5pt}
\textit{Acknowledgment.} We thank Chang Liu and Xiao Chen for helpful discussions. PZ is partly supported by the Walter Burke Institute for Theoretical Physics at Caltech. ZY is supported by the Key Area Research and Development Program of Guangdong Province (Grant No. 2019B030330001), the National Natural Science Foundation of China (Grant No.~12074440), and Guangdong Project (Grant No.~2017GC010613).

\textit{Note Added.} When completing this work, we noted that a scrambling transition was also revealed in a random unitary circuit that exchanges qubits with an environment \cite{Weinstein:2022yce}. The study focused on the evidence from the out-of-time-order correlator.

\bibliography{opensize20230527.bbl}

\end{document}